\documentclass[sigconf]{acmart}

\usepackage{balance}

\usepackage{xspace}
\usepackage{subcaption}
\usepackage{multirow}
\usepackage{enumitem}
\usepackage{listings}
\usepackage{makecell}
\usepackage{colortbl}

\makeatletter
\AtBeginDocument{%
  \begingroup
  \small
  \let\tmp@n@s\f@size
  \let\tmp@n@b\f@baselineskip
  \normalsize
  \let\tmp@s@s\f@size
  \let\tmp@s@b\f@baselineskip
  \xdef\semismall@size{\fpeval{(\tmp@n@s+\tmp@s@s)/2}}%
  \xdef\semismall@baselineskip{\fpeval{(\tmp@n@b+\tmp@s@b)/2}}%
  \endgroup
}
\newcommand{\semismall}{\fontsize{\semismall@size}{\semismall@baselineskip}\selectfont}

\newcommand{\toolName}{Task\-Au\-dit\xspace}
\newcommand{\Func}{Functiona11ity\xspace}
\newcommand{\func}{functiona11ity\xspace}

\newif\ifdisabletrackchanges
\disabletrackchangesfalse 
\disabletrackchangestrue 

\definecolor{darkpastelred}{rgb}{0.92, 0.2, 0.18}
\definecolor{burntorange}{rgb}{0.75, 0.38, 0.05}

\ifdisabletrackchanges
  \newcommand\del[1]{%
  }%
  \newcommand\add[1]{#1}
  \colorlet{AddColor}{black}
  
  \newcommand\addDiscuss[1]{}
  \newcommand\deleteDiscuss[1]{}
\else
  \usepackage[normalem]{ulem}
  \newcommand\del[1]{%
    \textcolor{darkpastelred}{\sout{#1}}%
  }
  \newcommand\add[1]{%
  \textcolor{Cerulean}{#1}%
  }
  \colorlet{AddColor}{Cerulean}
  
  \newcommand\addDiscuss[1]{\textcolor{burntorange}{\uuline{#1}}}
  \newcommand\deleteDiscuss[1]{\textcolor{burntorange}{\uuline{#1}}}
\fi

\newcommand\edit[2]{\del{%
#1%
}\add{#2}}

\newsavebox{\mycitebox}

\AtBeginDocument{%
  }

\setcopyright{acmlicensed}
\copyrightyear{2026}
\acmYear{2026}
\setcopyright{cc}
\setcctype{by}
\acmConference[CHI '26]{Proceedings of the 2026 CHI Conference on Human Factors in Computing Systems}{April 13--17, 2026}{Barcelona, Spain}
\acmBooktitle{Proceedings of the 2026 CHI Conference on Human Factors in Computing Systems (CHI '26), April 13--17, 2026, Barcelona, Spain}
\acmPrice{}
\acmDOI{10.1145/3772318.3791415}
\acmISBN{979-8-4007-2278-3/2026/04}

\begin{document}

\title[TaskAudit: Detecting Functiona11ity Errors in Mobile Apps via Agentic Task Execution]{\toolName: Detecting Functiona11ity Errors in Mobile Apps via Agentic Task Execution}


\author{Mingyuan Zhong}
\orcid{0000-0003-3184-759X}
\affiliation{%
  \department{Computer Science \& Engineering}
  \institution{University of Washington}
  \city{Seattle}
  \state{WA}
  \country{USA}
}
\email{myzhong@cs.washington.edu}

\author{Xia Chen}
\orcid{0009-0002-8918-5920}
\affiliation{%
  \institution{Carnegie Mellon University}
  \city{Pittsburgh}
  \state{PA}
  \country{USA}
}
\email{stanleyistool@gmail.com}

\author{Davin Win Kyi}
\orcid{0009-0006-8030-5024}
\affiliation{%
  \department{Computer Science \& Engineering}
  \institution{University of Washington}
  \city{Seattle}
  \state{WA}
  \country{USA}
}
\email{davin123@cs.washington.edu}

\author{Li Chen}
\orcid{0009-0002-6918-8120}
\affiliation{%
  \institution{Carnegie Mellon University}
  \city{Pittsburgh}
  \state{PA}
  \country{USA}
}
\email{meglichen23@gmail.com}

\author{James Fogarty}
\orcid{0000-0001-9194-934X}
\affiliation{%
  \department{Computer Science \& Engineering}
  \institution{University of Washington}
  \city{Seattle}
  \state{WA}
  \country{USA}
}
\email{jfogarty@cs.washington.edu}

\author{Jacob O. Wobbrock}
\orcid{0000-0003-3675-5491}
\affiliation{%
  \department{The Information School}
  \institution{University of Washington}
  \city{Seattle}
  \state{WA}
  \country{USA}
}
\email{wobbrock@uw.edu}


\begin{abstract}

Accessibility checkers are tools in support of accessible app development, and their use is encouraged by accessibility best practices. 
However, most current checkers evaluate static or mechanically-generated contexts, failing to capture common accessibility errors impacting mobile app functionality.
\edit{We present \toolName, an accessibility evaluation system that focuses on detecting \textit{functiona11ity} errors through simulated interactions.}
{In this work, we define \textit{functiona11ity} errors as accessibility barriers that only manifest through interaction}
\edit{[\textit{Functiona11ity} is}{(i.e., named according to} a blend of ``functionality'' and ``accessibility''\edit{ (a11y).]}{).}
We introduce \textit{\toolName}, which comprises three components: a Task Generator that constructs interactive tasks from app screens, a Task Executor that uses agents with a screen reader proxy to perform these tasks, and an Accessibility Analyzer that detects and reports accessibility errors by examining interaction traces. 
\add{Our} evaluation on real-world apps shows that \toolName detects 48 \func errors from 54 app screens, compared to between 4 and 20 with existing checkers. 
Our analysis demonstrates common error patterns that \toolName can detect in addition to \add{those from} prior work, including label-functionality mismatch, cluttered navigation, and inappropriate feedback.

\end{abstract}




\maketitle

\section{Introduction}
    Automated accessibility checkers are commonly used by mobile app designers, developers, and quality assurance testers to identify accessibility problems.
These tools are recommended by the Web Accessibility Initiative (WAI)~\cite{w3c_evaltools} and supported by platform owners~\cite{a11y_scanner, a11y_inspector} and various external organizations~\cite{axe, wave, caat_report}.

However, these checkers only perform static, heuristic assessments of a given screen's accessibility in an app.
As a result, automated checkers have limited error coverage, detecting only four out of 40 problem types identified in a study with blind and \edit{partially sighted}{low vision} participants~\cite{Carvalho_2018_a11yproblems}.
Similarly, they detected between 17\% and 31\% of accessibility errors as identified by experts according to a recent analysis~\cite{zhong2025screenaudit}.
To improve coverage, crawlers have been developed to support large language models (LLMs) in identifying additional error types not supported by current systems~\cite{zhong2025screenaudit}.
Generative agents have also been adopted \add{by AXNav} to automate accessibility testing in quality assurance~\cite{taeb24axnav}, but \add{the system} still uses heuristic-based error detection.

\begin{figure*}
  \includegraphics[width=0.88\textwidth]{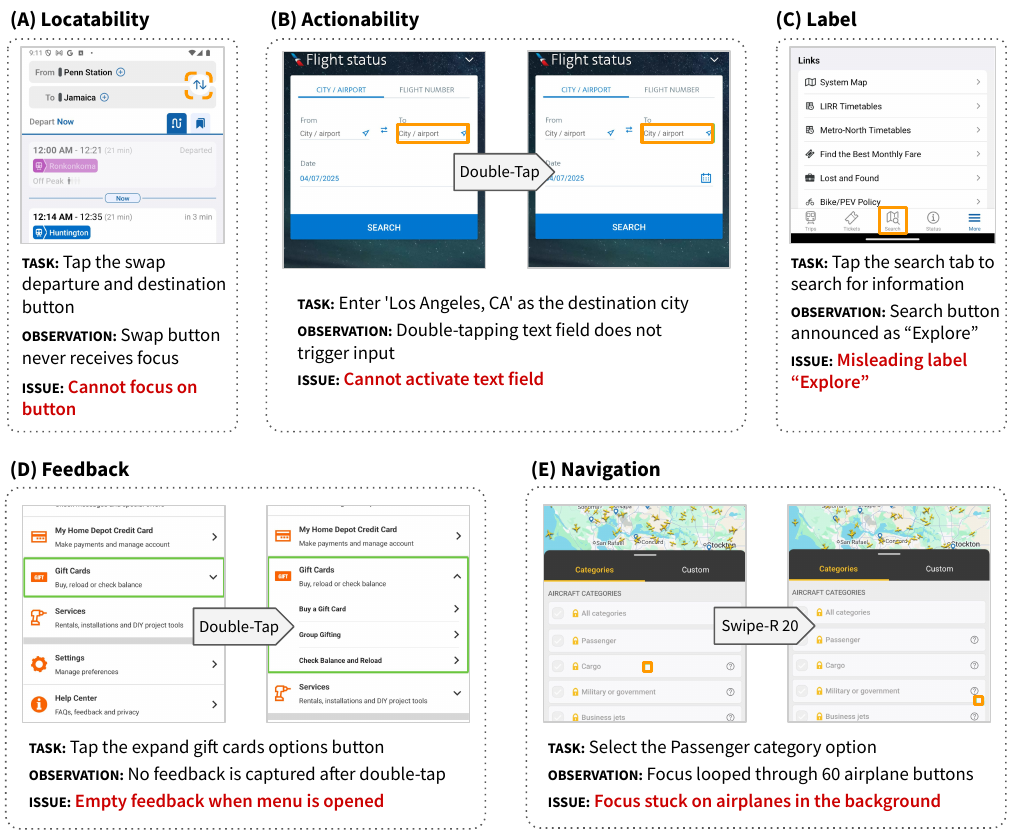}
  \caption{Exemplar errors from the five \func error categories that \toolName detected: \textit{Locatability}, \textit{Actionability}, \textit{Label}, \textit{Feedback}, and \textit{Navigation}. \add{Each example shows the task attempted (\textsc{task}), a summary of agent observation (\textsc{observation}), and its identified issue (\textsc{issue}). Screenshots show elements with focus (solid outline) or intended element (dashed outline). ``Swipe-R 20'' means 20 swipe-right gestures to traverse 20 elements.}}
  \label{fig:exp1}
  \Description{Five screenshots from mobile apps illustrate different functionality errors. Part (a) shows a locatability error where a button cannot be focused on. Part (b) shows an actionability error where a text field cannot be activated. Part (c) shows a label error where a search tab has a misleading label. Part (d) shows a feedback error where a menu gives no audible feedback when opened. Part (e) shows a navigation error where excessive swipes are needed to move past background elements.}
\end{figure*}

These advancements, while encouraging, overlook an important aspect \edit{in ensuring the accessibility of mobile apps}{of mobile app accessibility}: their functionality.
We define a \textit{\func error} as an accessibility barrier that only manifests through interaction, where the static state of the UI appears accessible, but the dynamic behavior fails to meet expectations or WCAG criteria~\cite{w3c_wcag2.2} (e.g., an action producing no feedback, a button's function not matching its label).
In this work, we focus on \func errors that manifest through screen readers.
Figure~\ref{fig:exp1} shows examples from five \func error categories we identified: \textit{Locatability}, \textit{Actionability}, \textit{Label}, \textit{Feedback}, and \textit{Navigation}.
We will introduce these in detail below.

Salehnamadi et al.~\cite{salehnamadi_latte_chi21, Salehnamadi2023groundhog} took initial steps toward addressing this gap with accessibility crawlers that detected two types of functionality issues: button locatability and actionability.
Although such systems represent a promising direction for detecting and addressing \func errors, 
we show that they are prone to generating false positives in real-world apps. 
In addition, their rule-based approach leaves out several types of \func errors.
In our opinion, effective detection of \func errors requires systems that not only adaptively explore applications via accessibility services, but also semantically interpret interface behavior.

As an exploration in expanding automated evaluation capabilities, we present \textit{\toolName}, an accessibility evaluation system that focuses on detecting \func errors through simulated interactions.
\toolName comprises three components: a Task Generator that constructs interactive tasks from app screens, a Task Executor that uses agents with a screen reader proxy to perform these tasks, and an Accessibility Analyzer that detects and reports accessibility errors by examining interaction traces.
Unlike Groundhog's mechanical crawling and direct pixel-wise comparison \add{approach}~\cite{Salehnamadi2023groundhog}, \toolName's agents attempt to understand the semantic goal and evaluate the outcome of an interaction sequence.
Compared to current agentic analyses such as AXNav~\cite{taeb24axnav}, \toolName is not limited to heuristic testing (i.e., loop and missing button detection), but leverages LLMs to directly manipulate and analyze the screen reader experience.
\toolName \del{also }does not use vision when operating under a screen reader to \edit{fully}{accurately} simulate that experience.
Table~\ref{tab:method_comparison} outlines the technical differences between these tools.

\begin{table*}[t]
    \small
    \centering
    \caption{Overview of mobile accessibility analysis techniques for interactive tasks. \add{\textit{\toolName} is the current tool compared to prior tools \textit{Groundhog} and \textit{AXNav}.}}
    \label{tab:method_comparison}
    \renewcommand{\arraystretch}{1.2}
    \begin{tabular}{l c c c}
        \toprule
        & \textbf{Groundhog}~\cite{Salehnamadi2023groundhog} & \textbf{AXNav}~\cite{taeb24axnav} & \textbf{TaskAudit} \\
        \midrule
        \rowcolor{gray!10}
        \textbf{Task Type} & Functionality-based & Workflow-based & Functionality-based \\
        \textbf{Task Creation} & Automatic, via view hierarchy & Manual & Automatic, via visual understanding \\
        \rowcolor{gray!10} 
        \textbf{Task Execution} & Mechanical crawling & Agentic execution w/ vision & Agentic execution w/o vision \\
        \textbf{Assessment} & Pixel-wise comparison & Heuristic analysis & LLM analysis \\
        \bottomrule
    \end{tabular}
\end{table*}

To evaluate \toolName's performance, we explored the following research questions:
\begin{enumerate}
    \item[\textbf{RQ1}:] Can interactive tasks be identified from an app screenshot?
    \item[\textbf{RQ2}:] Can a multi-agent system perform within-screen tasks successfully via a screen reader proxy?
    \item[\textbf{RQ3}:] Can \toolName detect \func errors effectively?
\end{enumerate}

We conducted three experiments, each addressing one research question with existing and new datasets.
Results show that the Task Generator covered 69.4\% of interactive tasks in a crowd-labeled dataset \add{(RQ1)}.
When no accessibility errors exist, our task-executing agents performed 96.0\% of within-screen tasks successfully via the screen reader proxy \add{(RQ2)}.
Evaluation on 54 unique screens from real-world apps shows that our strategy detected 48 out of 78 \func errors, compared to between 4 and 20 with existing checkers \add{(RQ3)}. 
Our analysis demonstrates common error patterns that \toolName can detect in addition to \add{those from} prior work, including label-functionality mismatch, cluttered navigation, and inappropriate feedback.

\edit{We}{This research} makes the following contributions:
\begin{enumerate}
    \item The \toolName system that automatically identifies interactive tasks on a mobile app screen, executes them via a screen reader proxy to test functionality, and analyzes accessibility errors by examining the execution traces.
    \item A performance evaluation demonstrating the feasibility of multi-agent task execution via a screen reader proxy and the system's ability to identify \func errors.
    \item A qualitative analysis of detected \func errors, such as label-functionality mismatch, cluttered navigation, and inappropriate feedback.
\end{enumerate}


\add{We envision \toolName as being part of the development cycle: it can be run when screens or interaction flows change alongside existing accessibility checks, adding the capability to execute interactions to detect \func errors.
\toolName can be integrated into accessibility inspection tools and IDEs, and optionally into continuous integration (CI) pipelines as a targeted check on high-impact flows, where developers can review and triage its reports to ensure accessibility before deployment.}

\section{Background: \Func Errors} \label{sec:background}
    We illustrate the concept of \func errors with examples in Figure~\ref{fig:exp1}.

\textbf{Locatability errors.} 
Borrowing from prior work~\cite{Salehnamadi2023groundhog}, one common \func error type is Locatability error.
For elements that can be invoked using touch interaction,
those that cannot be focused using a screen reader have Locatability errors.
Figure~\ref{fig:exp1}(a) shows one such example, where the swap direction button cannot be focused on.

\textbf{Actionability errors.}
Prior work~\cite{Salehnamadi2023groundhog} also proposed another error type: Actionability error.
Elements that can be focused but cannot be activated (i.e., clicked) by a screen reader have Actionability errors.
As shown in Figure~\ref{fig:exp1}(b), the input field does not respond to activation with a double-tap.

Locatability and Actionability errors are both covered by WCAG 2.1.1 ``Keyboard''\add{~\cite{w3c_wcag2.2}}, which requires all functionality to be operated through a keyboard-compatible interface (e.g., a keyboard, a screen reader, a switch device).

\textbf{Label errors.} 
We identify a Label error as a semantic mismatch between the element's label and its actual functionality.
Figure~\ref{fig:exp1}(c) shows one such example, where the ``search'' tab is incorrectly labeled as ``explore'' and thus can mislead a screen reader user who is looking for the search feature.
Label errors are covered under WCAG 4.1.2 ``Name, Role, Value'' and can be relevant to WCAG 1.1.1 ``Non-text Content'' depending on implementation\add{~\cite{w3c_wcag2.2}}.
While some Label errors may be detectable with static accessibility checking (e.g., a missing label), many potential issues cannot be definitively determined without examining functionality (e.g., a pictogram that may be interpreted in different ways), and therefore we included this category for completeness.

\textbf{Feedback errors.} This type of error relates to feedback quality after an action.
It occurs when actionable items, such as input fields, do not trigger an announcement after activation, or trigger feedback that is uninformative.
Figure~\ref{fig:exp1}(d) shows one such example, where activating a dropdown menu resulted in visual changes, but did not generate any audible feedback.
This can lead to confusion for screen reader users regarding the current state of the app.
These are covered by WCAG 3.2.2 ``On Input'' and WCAG 4.1.3 ``Status Messages''\add{~\cite{w3c_wcag2.2}}.
Compared to Label errors that concern announcement quality \textit{before} interactions, Feedback errors relate to quality \textit{after} interactions.

\textbf{Navigation errors.}
Navigation errors occur when the structure of interactive elements impedes efficient movement through the interface using a screen reader.
For example, in Figure~\ref{fig:exp1}(e), a large number of focusable airplane elements clutter the page, making it almost impossible to reach the main menu. 
This type of error \edit{overloads a screen reader's navigation order, leading}{can lead} to frustration and inefficiency during sequential navigation.
Navigation errors are addressed by WCAG 1.3.1 ``Info and Relationships,'' which requires content structure be programmatically determinable, and WCAG 2.4.1 ``Bypass Blocks,'' which requires mechanisms to skip repetitive content\add{~\cite{w3c_wcag2.2}}.

\section{Related Work}


Our research builds on related work on (1) manual accessibility inspection practices, (2) automatic accessibility evaluation techniques, and (3) LLM-driven task execution and its applications in accessibility evaluation.

\subsection{Accessibility Inspection Practices} \label{sec:practices}

The Web Content Accessibility Guidelines (WCAG) is an international standard for web content accessibility~\cite{w3c_wcag2.2}.
Although developed for the web, many of the WCAG requirements apply to mobile apps, and specific guidelines have been created for developers of mobile platforms~\cite{apple_guidelines, google_guidelines, microsoft_guidelines} and for the accessibility inspection of mobile apps~\cite{wild2023mobile, appt_guidelines, magentaa11y}.


In addition to guidelines, user testing is critical to ensure app accessibility.
In an assessment of WCAG 2.0, R{\o}men and Svan{\ae}s found that these guidelines only covered 32\% of accessibility problems experienced by users with disabilities~\cite{romen2012_wcag}.
In another study comparing the more recent WCAG 2.1, Mateus et al. found that users encountered 36 types of accessibility problems, while the guidelines only covered 22 (61\%)~\cite{mateus2020eval}.


User testing is not always feasible during accessibility assessments due to practical limitations.
Evaluation protocols, such as WCAG-EM~\cite{w3c_wcag_em} and EN 301 549~\cite{ETSI_EN_301549}, provide frameworks for systematically assessing websites and apps' accessibility.
However, their instructions were found to be nonspecific, especially when evaluating mobile apps~\cite{seixas2024exploring}.
In the same study, automated tools were also found to be limited in scope and availability. 
Evaluators therefore relied on manual examination, developing their own methodologies and using tools only for specific tasks such as checking color contrast.
Similarly, Pereira and Duarte reported that practitioners found automated checkers to be limited and relied on manual testing for accurate assessments~\cite{pereira2025evaluating}.

\subsection{Automated Accessibility Evaluation}
To support accessible app development and evaluation, a variety of automated accessibility evaluation tools have been created.
While these tools have been found to be under-utilized in current accessibility inspection practices~\cite{seixas2024exploring, pereira2025evaluating}, they have the potential to greatly improve the efficiency and coverage of accessibility testing.



Linters, such as those for Android Studio~\cite{accesslinter}, React Native~\cite{reactnative_eslint}, GitHub~\cite{accesslint}, and Deque's Axe Accessibility Linter~\cite{deque2024axelint} provide code-level accessibility checking for a limited set of issues during development. 
To evaluate accessibility in a more realistic setting and assess dynamically generated components, runtime accessibility checkers analyze a static snapshot of a webpage or an app screen to check for predefined errors.
Examples include Google's Accessibility Scanner~\cite{a11y_scanner}, Apple's Accessibility Inspector~\cite{a11y_inspector}, Deque's Axe~\cite{axe}, WAVE~\cite{wave}, and Google Chrome's Lighthouse~\cite{chromeLighthouse}.


To automate accessibility checking, crawlers programmatically invoke UI controls and expand the coverage of app states and functionalities.
MATE randomly activates elements and applies heuristic checking similar to runtime checkers~\cite{eler2018mate}.
\del{Recent large-scale mobile accessibility assessments adopt similar techniques~\cite{fok_large-scale_chi22, yan_current_state_accessibility, chen2022accessible}.}
To detect functionality errors, Groundhog~\cite{Salehnamadi2023groundhog} and BAGEL~\cite{chiou2023bagel} implement crawlers to simulate the activation of UI controls or navigation within a screen.
These automated checkers can overwhelm developers with redundant recommendations. Swearngin et al.~\cite{swearngin2024towards} addressed this by introducing a screen grouping model that de-duplicates and summarizes unique accessibility issues in apps.

\add{
Recent large-scale mobile accessibility assessments adopt crawling techniques to capture app snapshots for analyses~\cite{fok_large-scale_chi22, yan_current_state_accessibility, chen2022accessible}.
Beyond automated crawlers, Ross et al. adopt an epidemiological lens on mobile accessibility~\cite{ross_epidemiology_taccess20}. 
In particular, they examine image-based button labeling practices across thousands of Android apps~\cite{ross_examining_assets18} and measure the prevalence and distribution of accessibility problems in mobile ecosystems.
These studies characterize how accessibility errors emerge and persist at scale, and they motivate techniques that provide deeper analyses within individual apps beyond static analysis.
}


Another line of work expands accessibility evaluation coverage with AI.
López-Gil et al.~\cite{lopezgil2024web} explored automating certain WCAG success criteria using large language models (LLMs) that previously required manual checking.
HindDroid~\cite{liu2024Unblind} generates hints for inputs with an implication of supporting accessible app development and testing.
ScreenAudit~\cite{zhong2025screenaudit} crawls a screen for screen reader transcripts and generates an accessibility report based on an LLM analysis.
Recent work has also sought to provide code-level support, using LLMs, to detect and correct web accessibility errors including missing alt-text, improper headings, inadequate color contrast, and non-descriptive link descriptions \cite{delnevo2024interaction, huang2024access, othman2023fostering}.



Currently, automated tools remain limited in detecting functionality errors~\cite{mateus2020eval, seixas2024exploring, pereira2025evaluating}.
Our work combines crawler-based and AI-driven techniques to improve coverage of these errors.

\subsection{LLM-Driven Task Execution}

The core technology that enabled \toolName is automated task execution by LLM-driven agents, 
which are built upon recent advances in UI understanding and multi-agent systems.
Here, we introduce current UI understanding techniques, automated task execution, and their applications in accessibility evaluation.

\subsubsection{UI Understanding}

To acquire semantic understandings of a given UI, machine learning models have been developed to detect common UI components~\cite{zhang_screen_chi21, lu2024omniparser}, predict whether they are interactive~\cite{swearngin2019modeling, schoop2022predicting}, and generate captions for each component~\cite{li2020widget, lu2024omniparser}.
On the screen level, models such as Screen Parsing~\cite{wu2021screen} and Graph4GUI~\cite{jiang2024graph4gui} aim to recover structural information between UI components.
Screen2Vec~\cite{li2021screen2vec} enables robust screen comparisons and generalized task flow learning.
Screen2Words~\cite{wang2021screen2words} generates textual screen summaries using a multi-modal approach.
Recent advances in vision language models have led to OmniParser~\cite{lu2024omniparser} and ScreenAI~\cite{baechler2024screenai}, which combine many of the above capabilities to support screen annotation and question answering.

Much of the UI understanding work laid the foundation for automated task execution.
However, in our system, we rely solely on screen reader outputs for task execution, without using screenshots.
Instead, we leverage OmniParser~\cite{lu2024omniparser} to support the generation of interactive tasks to be evaluated by \toolName.

\subsubsection{Automated Task Execution}

Recent studies have utilized learning-based agents that leverage semantic UI understandings to automate task execution.
Agents rely on large language models (LLMs) to interpret interfaces and execute user instructions, such as navigating web and mobile interfaces~\cite{zhou2024webarena, huang2025prompt2task, song2024visiontasker, wen2024autodroid}. 
Other systems further integrate natural language prompts from users to guide task completion~\cite{vu2024gptvoicetasker}, or use pixel-based recognition techniques to infer UI semantics necessary for task execution~\cite{shaw2023pixels}. 
Multi-agent implementations distribute responsibilities across individual agent modules dedicated to planning, decision-making, and reflection or monitoring: Mobile-Agent-V2~\cite{wang2024mobileagentv2} separates planning, action, and reflection tasks among dedicated agents.
Prompt2Task~\cite{huang2025prompt2task} decomposes the task execution pipeline into specialized agents handling prompt analysis, information retrieval, task parsing, grounding to UI actions, and task assessment.
VisionTasker~\cite{song2024visiontasker} integrates UI understanding, task planning, and execution within modular sub-components of a single agent. 
Common themes across these approaches include the distribution of responsibilities such as semantic interpretation, interaction planning, and task execution.
In \toolName, we adopt a similar multi-agent design but adapt it to operate via a screen reader, without ``seeing'' the screen.

\subsubsection{Application in Accessibility Evaluation}

Traditional accessibility testing tools such as GroundHog~\cite{Salehnamadi2023groundhog} and FeedLack~\cite{ko2011feedlack} offer thorough identification of certain accessibility errors by mechanically crawling app states or analyzing source code to detect missing user interface feedback.
Without semantic understanding, they lack coverage of nuanced functionality issues.
AXNav~\cite{taeb24axnav} employs multiple agents (planner, action, evaluation) that execute pre-defined tasks to assess UI accessibility.
However, AXNav operates the app using conventional controls and does not directly evaluate the user experience provided by accessibility services.
ScreenAudit~\cite{zhong2025screenaudit} uses an LLM to analyze TalkBack transcripts with screen metadata, but primarily relies on mechanical crawling without directly testing functionality.



Building upon multi-agent task execution approaches~\cite{wang2024mobileagentv2, taeb24axnav} and LLM-driven accessibility checkers~\cite{lopezgil2024web, liu2024Unblind, zhong2025screenaudit}, we leverage LLM-driven task execution agents to directly evaluate functionality through a screen reader proxy. 
Unlike existing tools that mechanically crawl interfaces or examine metadata~\cite{Salehnamadi2023groundhog, ko2011feedlack, zhong2025screenaudit}, our approach targets semantic-dependent interactions, surfacing errors tied to the actual screen reader user experience.

\section{The \toolName System}
    \begin{figure*}[t]
  \includegraphics[width=0.85\textwidth]{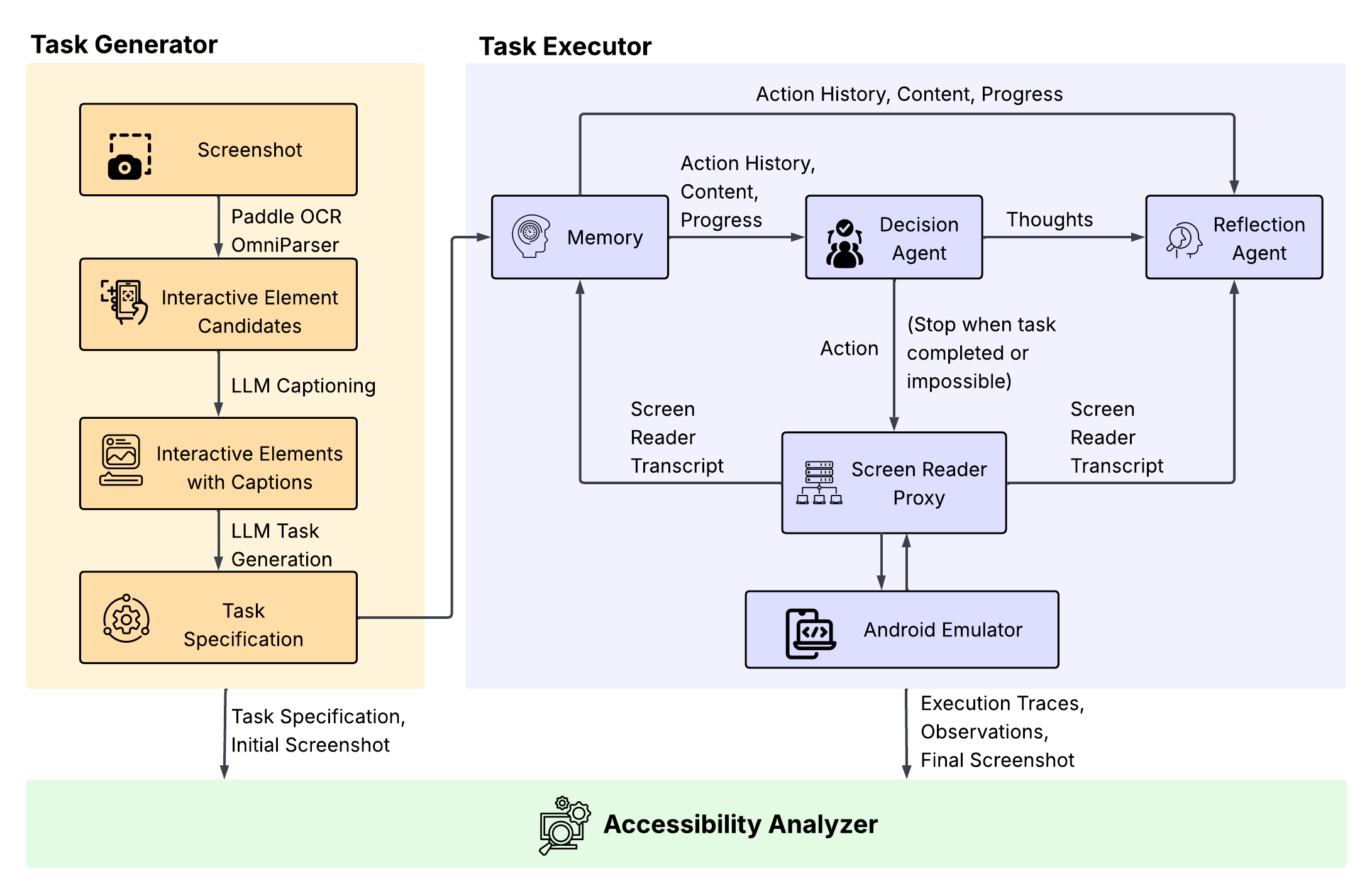}
  \caption{\toolName's system architecture, comprising a Task Generator, Task Executor, and Accessibility Analyzer. The Task Generator produces task specifications from an initial screenshot. The Task Executor accepts a task and utilizes a multi-agent approach to execute it via a screen reader proxy on an Android emulator. Execution traces and related metadata are sent to the Accessibility Analyzer for evaluation.}
  \label{fig:system}
  \Description{System architecture diagram for TaskAudit. A 'Task Generator' block takes a screenshot, uses OCR and LLMs to identify interactive elements and generate task specifications. A 'Task Executor' block shows a multi-agent system (Decision, Reflection, Memory) interacting with an Android Emulator via a Screen Reader Proxy to execute the task, producing traces. An 'Accessibility Analyzer' block takes inputs from the generator and executor to report errors. Arrows indicate data flow between components.}
\end{figure*}

Informed by our understanding of \func errors,
we created \toolName to explore detecting such errors in mobile apps through automated task execution.

As shown in Figure~\ref{fig:system}, the \toolName system consists of three components:
a \textit{Task Generator} that constructs interactive tasks from a mobile app screen,
a \textit{Task Executor} that performs each task through a screen reader proxy using custom task-executing agents,
and an \textit{Accessibility Analyzer} that examines collected task execution traces to determine and report any errors.
We adopted OpenAI's GPT-4o~\cite{gpt4o} for \toolName, but expect the system design to be adaptable to other language models.

\subsection{Task Generator}
The Task Generator constructs potential tasks from a screenshot captured on the screen under evaluation.
As no consistent guidelines currently exist for mobile accessibility evaluation~\cite{pereira2024exploring}, we reference the Website Accessibility Conformance Evaluation Methodology (WCAG-EM) 1.0 document published by the W3C~\cite{w3c_wcag_em}.
In particular, Requirement 4.b on accessibility auditing is the most relevant and involves checking ``functionality, entering data, notifications, and other interaction.''
Therefore, we designed the Task Generator to generate a \textit{task specification} for each interactive element appearing in the screenshot.
We chose a screenshot-based approach that allows the system to identify interactive elements that are rendered visually but not properly exposed in the accessibility tree, as we will demonstrate through Groundhog's results in Section~\ref{sec:func-error} and illustrated in Figure~\ref{fig:gh_fail}.
This also allows the Task Generator to function where direct access to an app's view hierarchy is not available, reflecting many real-world quality assurance workflows (e.g., an app may hide or obfuscate the view hierarchy or use a framework that does not properly expose metadata).

\begin{figure*}[t]
  \includegraphics[width=0.75\textwidth]{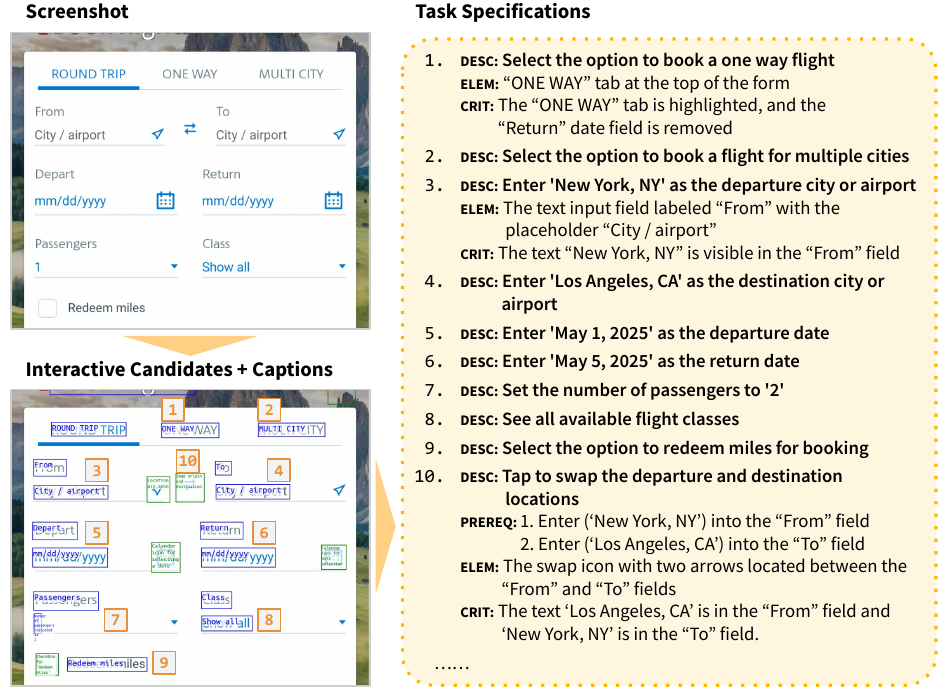}
  \caption{\toolName's Task Generator parses the interactive elements from a screenshot and generates task specifications. For brevity, we only present selected examples from the specifications, including \textsc{desc} for description, \textsc{prereq} for prerequisites, \textsc{elem} for element, and \textsc{crit} for success criterion.}
  \label{fig:task_gen}
  \Description{Task Generator's process for creating tasks. On the left, a screenshot of a flight booking app is analyzed to identify interactive elements, which are outlined with numbered boxes. On the right, a list of corresponding task specifications is shown, such as "Select the option to book a one way flight" and "Enter 'New York, NY' as the departure city or airport."}
\end{figure*}

To ensure reliable and consistent task generation, the Task Generator adopts a three-step process, with the first two steps in line with recent work in UI understanding~\cite{baechler2024screenai, lu2024omniparser}.
Figure~\ref{fig:task_gen} illustrates the process with an example from an app that we examined.
\begin{enumerate}
    \item \textit{Identifying interactive element candidates}
        from a screenshot using Paddle OCR~\cite{li2022ppocrv3} for text extraction and OmniParser's icon detection model~\cite{lu2024omniparser}.
    \item \textit{Captioning.}
        Candidates are individually cropped from the screenshot with surrounding context and sent to GPT-4o for captioning. Specifically, we prompt the model to output a descriptive name and element category (i.e., \textit{information}, \textit{action}, \textit{input}, and \textit{navigation}) for each element.
        These categories are derived from the WCAG-EM 1.0~\cite{w3c_wcag_em}, and Apple~\cite{apple_components} and Google's~\cite{google_components} classifications of UI components.
        Figure~\ref{fig:task_gen} includes the descriptive names overlaid over their respective elements.
    \item \textit{Task specification generation.}
        The Task Generator filters out informational elements from the candidate list.
        The resulting list of interactive elements and an annotated screenshot (i.e., each element is indexed and highlighted in its bounding box) are sent to GPT-4o to generate specifications of interactive tasks.
        For each task, the model is prompted to generate a description with necessary details (e.g., time, location, name), identify a corresponding element and any prerequisites, and propose a task success criterion for the Task Executor and the Accessibility Analyzer to determine if a task is successful.
\end{enumerate}

\subsection{Task Executor}
Given a task description from the Task Generator, the Task Executor attempts to perform that task through a \textit{screen reader proxy}, using a multi-agent execution workflow.
The goal is to simulate how a person using screen reader would interact with the app, and collect screen reader output for accessibility evaluation.

\begin{figure*}[t]
  \includegraphics[width=0.85\textwidth]{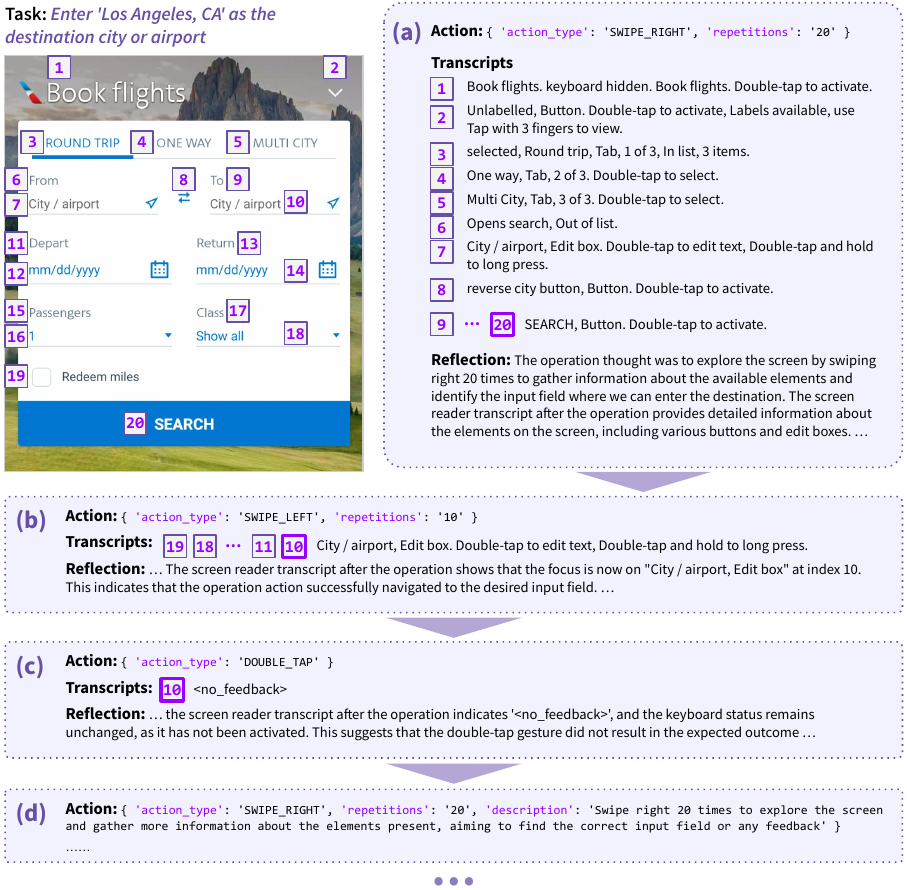}
  \caption{\toolName's Task Executor executes a task which ultimately leads to a \func error. In each step, the Decision agent generates an action, transcripts are collected through the screen reader proxy, and the Reflection agent assesses the outcome. Throughout the process, the accessibility focus changes. The last focused element in each step is highlighted in the transcripts. This example includes an input control that does not respond to activation (Step c), which led the Decision agent to attempt other pathways (Step d). However, this proved to be impossible and the execution ultimately terminated after a loop was detected (not shown for brevity).}
  \label{fig:agent_exec}
\end{figure*}

\subsubsection{Screen Reader Proxy}
A screen reader serves as a \textit{proxy} between an app's original interface and the manifest interface that a blind person perceives~\cite{zhang_interaction_chi17}, converting a graphical user interface to a mostly linear interface announced by voice.
In our system, the screen reader proxy \edit{serves as a link between}{links} the task-executing agents and the \textit{environment}, i.e., the app interface manifested through the screen reader.
\edit{In order to efficiently capture screen reader output, we adopted a modified version of Android's TalkBack as described in prior work~\cite{zhong2025screenaudit}, which allowed the direct capture of its transcripts from the internal pipeline before they are verbally announced.}{We adopt a modified version of Android's TalkBack~\cite{zhong2025screenaudit} that allows direct capture of transcripts from its internal pipeline before they are announced.}

\del{We support a subset of possible TalkBack interactions~\cite{talkback}: linear navigation with left or right swipes (i.e., consistent with keyboard navigation using the Tab key) and element activation (i.e., clicking) with double-taps.}
The screen reader proxy accepts one or multiple swipe gestures \add{(for navigating between elements)}, or a single double-tap gesture \add{(for clicking on an element)} in each interaction.
These gestures are \del{also }consistent with other screen readers, such as VoiceOver~\cite{voiceover}.
We also support standard Android gestures, including going back and text entry.
The gestures are invoked using our modified TalkBack service on an Android emulator, and \edit{the resulting transcripts are collected.}
{the screen reader proxy exports the resulting transcripts using the Android logging system, which are then forwarded to the Task Executor via a socket connection.
Other data, such as screenshots and accessibility metadata, are collected separately and saved for later analysis.
}
We did not support touch exploration in line with prior work~\cite{salehnamadi_latte_chi21, Salehnamadi2023groundhog, zhong2025screenaudit} due to a lack of behavioral data on touch exploration from screen reader users, accessibility guidelines requiring full keyboard navigation support~\cite{w3c_wcag2.2}, and the exploratory nature of our work.
We recognize this as an opportunity for future research.

\subsubsection{\edit{Agent Scaffolding}{Multi-Agent Execution Workflow}}
\edit{We built a multi-agent system supported by LLMs to execute tasks proposed by the Task Generator.}{We implement the Task Executor as a multi-agent system supported by LLMs.}
Inspired by recent task-executing agents~\cite{huang2025prompt2task,wang2024mobileagentv2,song2024visiontasker}, we use distinct Decision and Reflection agents so the system can explicitly assess each action's outcome before planning the next step, which is particularly important because the agents do not see screenshots.
\edit{Our system follows a step-by-step task execution design.}{At each step, the Decision agent proposes an action, the proxy executes it and returns transcripts, and the Reflection agent evaluates the outcome, updating a shared Memory store.}
\del{The intent is to provide the multi-agent system with the exact set of information that a screen reader user would perceive.}
Figure~\ref{fig:agent_exec} illustrates this loop with a task execution example.
\del{The system has three main components: a \textit{Memory} mechanism, a \textit{Decision agent}, and a \textit{Reflection agent}.}
Prompts are available in Appendix~\ref{sec:appendix-prompt}.

\textbf{Memory.}
The Memory mechanism stores task-relevant information of the user's goal, environment setup, and execution history of previous steps to help execute the task. \del{At each step, it updates its memory dynamically as the task progresses, based on the new observation after the performed action, in order to provide context when planning the next action.}

\textbf{Decision Agent.}
The Decision agent generates actions based on the task description, current progress, transcripts, and prior reflections.
It operates within a predefined \textit{operation space} of four environment actions (swipe, double-tap, back, type) and three meta-actions (wait, mark task as complete, mark task as impossible).

\textbf{Reflection Agent.}
The Reflection agent assesses whether each action's outcome matches expectations given the prior state and transcripts and, when it fails, identifies likely reasons.
This reflection process follows prior work on task-executing agents~\cite{wang2024mobileagentv2}.
Reflections are stored step-by-step to support later analysis but are not themselves treated as accessibility judgments.

\subsubsection{\edit{Implementation}{Exploration and Navigation} Details}
\edit{We implement the Task Executor on top of an Android emulator to ensure task execution consistency.
For each screen under analysis, an initial screenshot and an Android Virtual Device (AVD) snapshot~\cite{android_snapshots} is captured.
The snapshot is restored prior to the execution of each task using standard adb commands.}{We run the Task Executor on top of Android emulator snapshots~\cite{android_snapshots} and, for each screen under analysis, capture and restore an Android Virtual Device (AVD) snapshot before executing each task to ensure consistent execution.}

\edit{A screen reader typically does not produce announcements without changes or interactions on screen.
When presented with a new task in its initial, unexplored state, the decision agent does not have contextual information regarding the screen, unlike traditional task-executing agents that observe screenshots or UI hierarchies.
Therefore, the agent must explore the environment and identify relevant UI elements.
To facilitate this, we explicitly prompt the decision agent to perform $k$ ``swipe right'' actions until it detects an element related to the task.
In our implementation, we set $k=20$ and permit up to three initial exploration attempts, allowing a maximum of 60 elements to be explored.}{Because a screen typically does not produce transcripts until interacted with, the Decision agent must first explore the environment and identify relevant elements, so we explicitly prompt it to perform up to $k=20$ ``swipe right'' actions (up to three attempts) until it detects an element related to the task, allowing a maximum of 60 elements to be explored.}

In some cases, \edit{after the decision agent has found its target element, we observed that the swipes needed to get back to the elements may not be consistent with the number of elements announced.
This is usually a result of TalkBack's logic of handling list navigation, which can cause confusion with the agent and result in more steps.}{TalkBack's list-handling logic makes the number of swipes needed to revisit elements inconsistent with the number of announced elements, which can confuse the agent.}
We mitigate this by allowing the agent to specify an object in the original transcript, and the screen reader proxy programmatically navigates until that element is matching based on string similarity.





\begin{figure*}[t]
  \includegraphics[width=0.98\textwidth]{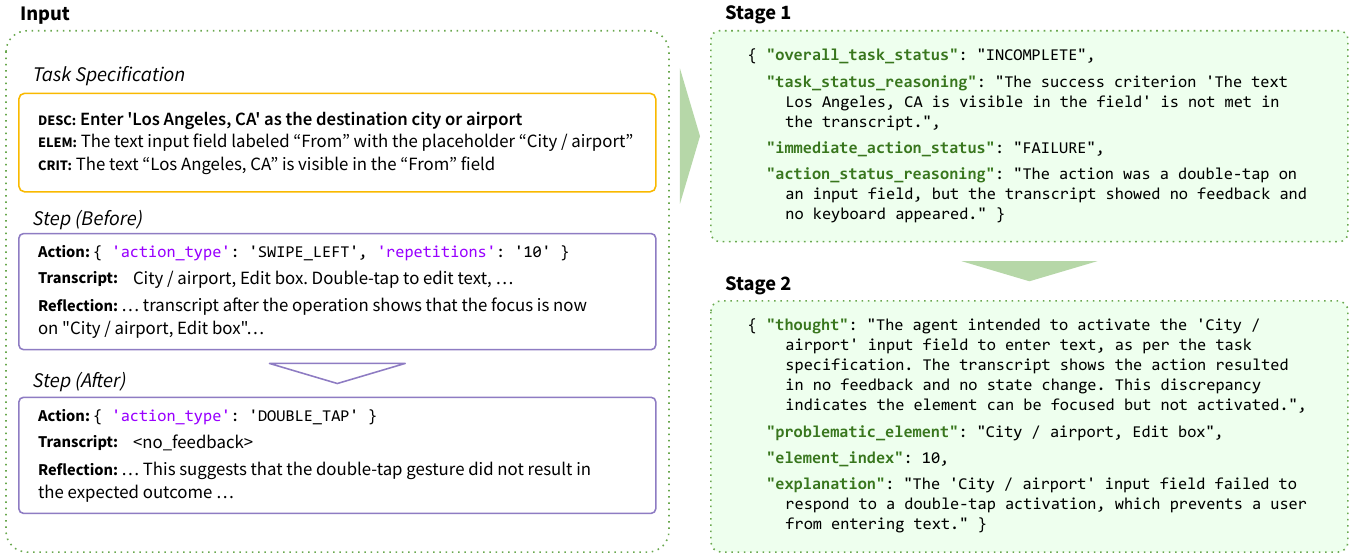}
  \caption{\toolName's Accessibility Analyzer processes one step in the execution trace from Figure~\ref{fig:agent_exec} [(b) to (c)]. It uses a two-stage prompting process to identify the root cause and reports an error.}
  \label{fig:a11y_analyzer}
  \Description{The two-stage process of the Accessibility Analyzer. The input shows an agent's action that resulted in "no feedback." In Stage 1, the system determines the immediate action failed. In Stage 2, it performs a root cause analysis and concludes that "The 'City / airport' input field failed to respond to a double-tap activation."}
\end{figure*}

\subsection{Accessibility Analyzer}
The Accessibility Analyzer processes task execution traces to interpret the outcome of each interaction and identify possible \func errors.
The analyzer performs a step-by-step evaluation of the execution trace.
Because a task may be completed before all planned actions are executed, the overall Task Specification is checked after every action.
Central to the analysis is the execution trace collected during task execution.
Each trace includes a list of executed steps which contains the following information corresponding to each action performed via the screen reader proxy:
\begin{enumerate}
    \item Summary of execution progress before each action.
    \item Thoughts and reasoning behind each action proposed.
    \item Observation (screen reader transcripts) after each action.
    \item Reflection of whether each action met expectation.
\end{enumerate}

The analysis is conducted in a two-stage process for each step in the trace, which we illustrate with an example in Figure~\ref{fig:a11y_analyzer}.
Prompts are available in Appendix~\ref{sec:appendix-analyzer}.

First, for each action performed, we use GPT-4o to determine its outcome.
The model's goal is to check if the overall task's success criterion has been met during any of the steps, allowing for early task completion.
The model also validate the immediate action's accessibility using a set of action-specific heuristics.
These heuristics include verifying that a keyboard appears after tapping a text field, that the screen content changes after a scroll action, and that new elements are announced after activating a menu.

Second, if an action is marked as a FAILURE, its trace and the original Task Specification are further analyzed using a specialized prompt.
This model uses a structured, Chain-of-Thought style reasoning process to determine if the failure was caused by accessibility errors in the app, or that no accessibility error has occurred if the failure is attributable to other factors, such as an agent execution error or a change in the app's state.
This process ensures that every action is validated against both its immediate expected outcome and the task's final goal, and that failures are reviewed before being flagged as accessibility errors.

\section{Performance Evaluation} \label{sec:perf-eval}
    \toolName adapts and combines many components towards implementing automated assessments of \func errors.
While some of these techniques have been published and evaluated in prior work, here we systematically evaluate them in the context of detecting \func errors.

We conducted experiments to assess \toolName's performance in three respects:
(1)~its ability of identifying tasks on a screen,
(2)~its ability of executing tasks through a screen reader proxy, when no accessibility errors are involved,
and (3)~its ability of detecting \func errors.
These first two assessments correspond to the Task Generator and the Task Executor respectively and the third experiment evaluates the entire \toolName pipeline.

To support these experiments, we adapted existing datasets and created our own datasets as needed.
We report results in similar tasks from prior work where available.

\subsection{Task Identification}
\subsubsection{Method}
We evaluate \toolName's task identification performance by comparing the interactive elements identified by the Task Generator against human annotated datasets from prior work~\cite{schoop2022predicting, li2020widget}.
Specifically, we ran the Task Generator over a human-annotated screenshot dataset containing UI element tappability and captions.
An element was considered detected if over 80\% of the bounding box extracted by the Task Generator overlaps with a tappable element from the dataset.
Two authors independently compared system-generated with crowdsourced captions, flagging captions as inconsistent if they were off-topic or had missing information.
The authors discussed any disagreements until they reached a consensus.

We did not compare with GroundHog~\cite{Salehnamadi2023groundhog} or other accessibility checkers in this experiment, because they require analyses of \textit{running} apps which we cannot perform on our annotated dataset.
However, we provide a detailed comparison of existing techniques in Section~\ref{sec:func-error}.

\subsubsection{Datasets}
We used three datasets in this experiment:
\begin{enumerate}
    \item RICO~\cite{deka_rico_uist17}, a large-scale dataset of mobile app screenshots.
    \item A crowdsourced \textit{UI tappability} dataset~\cite{schoop2022predicting}, where each UI element is labeled ``tappable'' or ``not tappable.''
    \item A crowdsourced \textit{widget captioning} dataset~\cite{li2020widget}, where each UI element has a caption about its functionality.
\end{enumerate}

The UI tappability and widget captioning datasets are random subsets of the RICO dataset.
The intersection of the two datasets contained 2,041 UI elements from 1,086 screens.
Each element in the widget captioning dataset had two or three human-generated captions.
Two of the authors independently reviewed and manually removed 815 elements whose captions appear to describe different functionalities (e.g., one is ``filter results'' and the other is ``menu bar'').
These inconsistencies are likely due to screenshot inconsistencies or crowdworker errors from the original dataset.
The final dataset contained 1,226 unique elements from 813 screens.

\subsubsection{Results}
Our Task Generator detected 911 of the 1,226 tappable UI elements (74.3\%).
Among these, 69.5\% of all buttons (643 of 925), 73.0\% of selection inputs (46 of 63), and 93.3\% of text inputs (222 of 238) were detected.
We did not calculate precision, as no available dataset has complete labels for all tappable elements.
False positives will have a limited impact on the final assessment, as the Task Executor identifies such elements as non-actionable during its run.
As a reference, Zhang et al.~\cite{zhang_screen_chi21} developed a UI element detection model that achieved a mean Average Precision (mAP) of 71.3\% on a proprietary dataset.
For generated captions, 851 (93.4\%) were consistent with the crowdsourced captions.
The overall actionable element identification rate was 69.4\% (851 of 1,226).

\subsection{Task Execution via Screen Reader Proxy}
\subsubsection{Method}
We measured the task execution success rate via screen reader proxy by generating and executing tasks on apps known to be accessible.
Our test set included apps with no known missing-label errors from Fok et al.~\cite{fok_large-scale_chi22} and all Google-developed apps from the Android in the Wild dataset~\cite{rawles2023aitw}, excluding Chrome.
From this set, we manually examined 26 apps to identify 36 screens with no apparent \func errors.

We ran our Task Generator on these screens to produce and execute 299 unique tasks via the screen reader proxy.
A task was considered successful if the agent correctly focused on a non-actionable element, activated an actionable element, or entered data into an input field.
Since each task only involved interacting with or altering the state of one element, we expected the task success rate to be high.

\subsubsection{Results}
Our agents successfully executed 287 of the 299 tasks, achieving a 96.0\% success rate.
Most failures (7 tasks, 2.3\%) were caused by the LLM generating incomplete task descriptions, such as specifying a search action without providing a query text.
The remaining 5 failures (1.7\%) were due to content refreshing on the screen between task generation and execution, a delay our system mitigates but cannot entirely eliminate.
The high success rate on these accessible screens demonstrates the robustness of our agentic task execution.

\subsection{\Func Error Detection} \label{sec:func-error}

\begin{table}
    \small
    \centering
    \caption{Performance comparison of accessibility checkers in detecting \func errors.}
    \label{tab:checker_performance}
    \renewcommand{\arraystretch}{1.2}
    \begin{tabular}{p{4cm} c c c}
        \toprule
        \textbf{Tool} & \textbf{Precision} & \textbf{Recall} & \textbf{F1} \\
        \midrule
        \rowcolor{gray!10}Accessibility Scanner & - * & 0.052 & - \\
        ScreenAudit & - * & 0.130 & - \\
        \rowcolor{gray!10}Groundhog & 0.142 & 0.256 & 0.183 \\
        TaskAudit & 0.676 & 0.615 & 0.644 \\
        \bottomrule
    \end{tabular}
    
    \bigskip
    \footnotesize \parbox{8cm}{ *\; We did not calculate precision for Accessibility Scanner and ScreenAudit because (1) they are not designed for detecting \func errors and that (2) they reported many other aspects of accessibility issues not covered by our dataset. Separating these from \func errors will inherently be subjective and we defer to related work~\cite{zhong2025screenaudit} for a more detailed comparison on these tools.}
\end{table}

\begin{table*}
    \small
    \centering
    \caption{Breakdown of causes for incorrect outcomes for Groundhog and TaskAudit.}
    \label{tab:outcome_breakdown}
    \renewcommand{\arraystretch}{1.2}
    \begin{tabular}{l l p{2.6cm} p{5.6cm} c}
        \toprule
        \multirow{2}{*}{\textbf{Tool}} & \multirow{2}{*}{\textbf{Outcome}} & \multirow{2}{*}{\makecell[l]{\textbf{System Component} \\ \textbf{of Error Origin}}} & \multirow{2}{*}{\textbf{Cause}} & \multirow{2}{*}{\textbf{Count (\%)}} \\
        \\
        \midrule
        
        \multirow{8}{*}{\textbf{Groundhog}} 
        & \multirow{5}{*}{\makecell[l]{False \\ Positive}} & \multirow{2}{*}{\makecell[l]{Locatability \\ Detection}} & \hspace{0.3em}$\dashrightarrow$ Hidden element & 27 (22.9\%) \\
        & & & \hspace{0.3em}$\dashrightarrow$ Container not intended to be focusable & 19 (16.1\%) \\
        \cmidrule(l){3-5}
        & & \multirow{2}{*}{\makecell[l]{Actionability \\ Detection}} & \hspace{0.3em}$\dashrightarrow$ Destination screen comparison issue & 50 (42.4\%) \\
        & & & \hspace{0.3em}$\dashrightarrow$ Container not intended to be actionable & 22 (18.6\%) \\
        \cmidrule(l){3-5}
        & & \multicolumn{2}{l}{\textit{Total False Positives}} & \textit{118, FPR: N/A} \\
        \cmidrule(l){2-5}
        & \multirow{2}{*}{\makecell[l]{False \\ Negative}} & \multicolumn{2}{l}{\textit{Locatability and Actionability Detection only}} & \textit{17, FNR': 45.9\%} \\
        \cmidrule(l){3-5}
        & & \multicolumn{2}{l}{\textit{Total False Negatives under all categories}} & \textit{58, FNR: 74.4\%} \\
        \midrule
        
        \multirow{11}{*}{\textbf{TaskAudit}} 
        & \multirow{6}{*}{\makecell[l]{False \\ Positive}} & \multirow{3}{*}{Agent Execution} & \hspace{0.3em}$\dashrightarrow$ Inaccurate navigation & 4 (17.4\%) \\
        & & & \hspace{0.3em}$\dashrightarrow$ Verbose or confusing transcripts & 2 (8.7\%) \\
        & & & \hspace{0.3em}$\dashrightarrow$ Screen reader proxy unreliable & 1 (4.3\%) \\
        \cmidrule(l){3-5}
        & & \multirow{2}{*}{Task Generator} & \hspace{0.3em}$\dashrightarrow$ Visual UI parsing error & 12 (52.2\%) \\
        & & & \hspace{0.3em}$\dashrightarrow$ Incorrect contextual understanding & 4 (17.4\%) \\
        \cmidrule(l){3-5}
        & & \multicolumn{2}{l}{\textit{Total False Positives}} & \textit{23, FPR: 9.8\%} \\
        \cmidrule(l){2-5}
        & \multirow{4}{*}{\makecell[l]{False \\ Negative}} & \multirow{2}{*}{\makecell[l]{Accessibility \\ Analyzer}} & \hspace{0.3em}$\dashrightarrow$ Did not identify error & 6 (20\%) \\
        & & & \hspace{0.3em}$\dashrightarrow$ Used alternative, accessible interaction & 3 (10\%) \\
        \cmidrule(l){3-5}
        & & Task Generator & \hspace{0.3em}$\dashrightarrow$ Did not generate task & 21 (70\%) \\
        \cmidrule(l){3-5}
        & & \multicolumn{2}{l}{\textit{Total False Negatives}} & \textit{30, FNR: 38.5\%} \\
        \bottomrule
    \end{tabular}
\end{table*}

\begin{figure*}
  \includegraphics[width=0.8\textwidth]{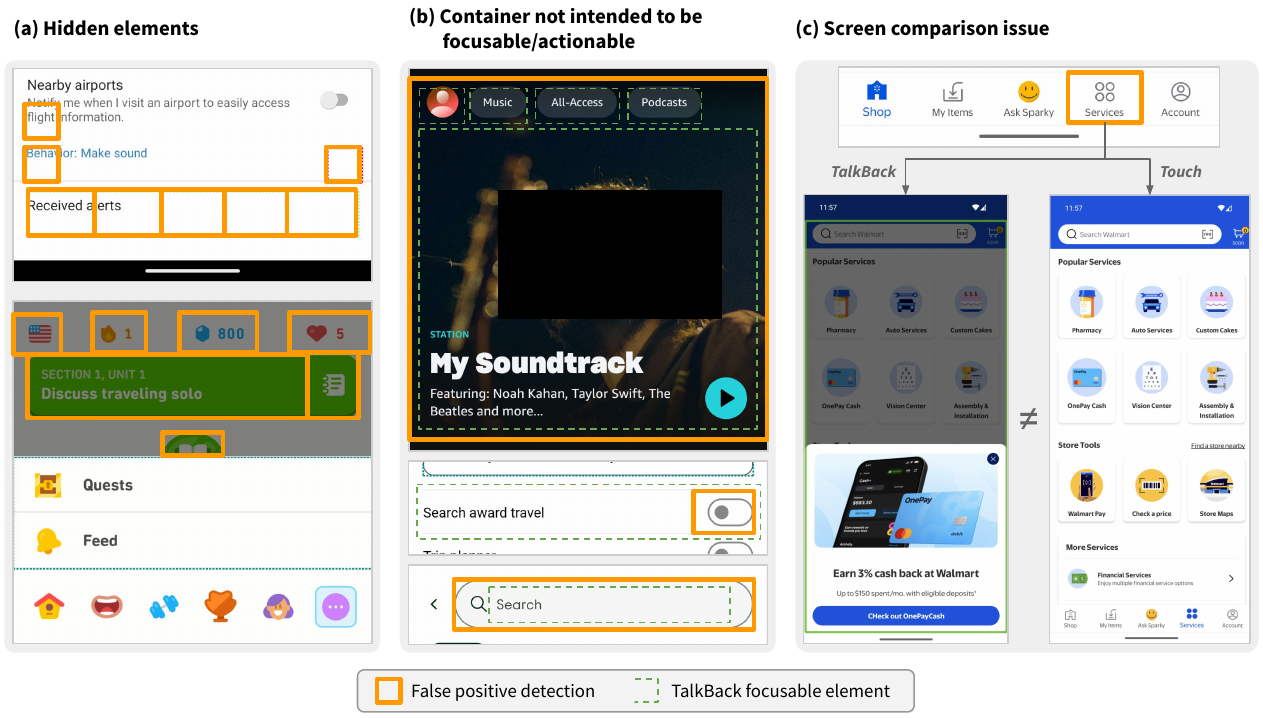}
    \caption{Examples of Groundhog's false positive detections highlighted in orange. (a) Elements covered by an overlay are incorrectly flagged. (b) Containers are incorrectly flagged when alternative elements provide full functionality and are focusable. Making the containers focusable would instead cause duplicates. (c) An Actionability error being incorrectly flagged because of a pop-up ad captured during screen comparison. }
  \label{fig:gh_fail}
  \Description{Three screenshots illustrate common causes for Groundhog's false positive errors. Part (a) shows hidden elements under an overlay that are incorrectly flagged. Part (b) shows a large, non-interactive container being flagged instead of the smaller, interactive elements within it. Part (c) shows how a pop-up ad can interfere with screen comparison, leading to an incorrect error report.}
\end{figure*}

\begin{figure*}
  \includegraphics[width=0.72\textwidth]{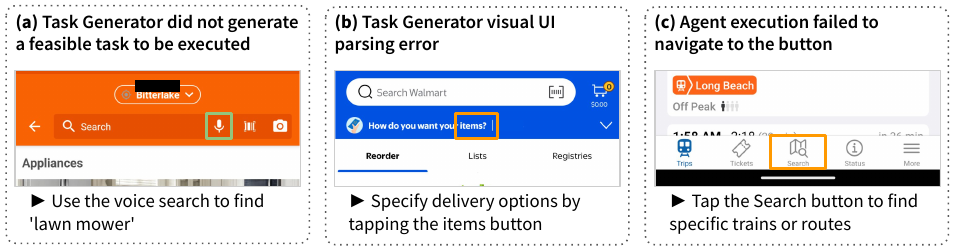}
    \caption{Examples of \toolName's false positive detections with corresponding task descriptions below. (a) Task Generator created an impossible task as the Executor has no speech capability. (b) Task Generator made an visual parsing error. (c) The Executor failed to navigate to an accessible button. }
  \label{fig:ta_fail}
  \Description{Three examples show the causes of TaskAudit's false positives. Part (a) shows an impossible task generated by the system to "Use the voice search." Part (b) shows a visual parsing error where the system misidentified a button. Part (c) shows a case where the agent failed to navigate to an accessible button.}
\end{figure*}

\subsubsection{Method}
We evaluated whether \toolName is capable of detecting and reporting \func errors and compared its performance against three state-of-the-art accessibility checkers: the Android Accessibility Scanner~\cite{a11y_scanner},\del{ a replicated version of} ScreenAudit~\cite{zhong2025screenaudit}, and Groundhog~\cite{Salehnamadi2023groundhog}.
We evaluated them on 54 unique screens from 14 apps.
Three of the authors independently examined all generated reports against the observations recorded in our \func dataset, recording their outcomes.

\subsubsection{Dataset}
We created a \func error dataset of 78 error instances, building on top of a recently published accessibility error dataset that included 14 apps with one unique screen per app~\cite{zhong2025screenaudit}.
Their dataset included 12 instances related to ``functionality,'' covering elements' keyboard access, focus, input, and feedback.
To cover a more diverse set of screens and functionalities, we included one additional screen under each navigational tab from each app.
\add{We included these additional screens because (1) top-level tabs are usually on critical paths of app functionality, and (2) they cover a diverse set of interactive components, such as buttons, text inputs, drop-downs, and selection elements.}
In total, \edit{this resulted in}{our dataset contained} 54 unique screens.
\add{As these apps were randomly selected according to prior work~\cite{zhong2025screenaudit}, we expect them to be representative of typical mobile apps.}

Three of the authors independently interacted with informational and interactive elements on all screens using both direct touch and Android TalkBack.
Following WCAG-EM~\cite{w3c_wcag_em}, the authors recorded observed behaviors and any accessibility issues.
\add{The authors conducted the interactions on an app outside of the selected apps to align their inspection methodologies before starting the annotation process.
No automated tools were used, except for TalkBack to simulate screen reader interactions.
}
We included interactions that received at least two annotations in the dataset.
In total, we examined 475 unique interactions and identified 78 interactions containing accessibility errors.
We classified these errors into five general categories: \textit{Locatability} (25), \textit{Actionability} (12), \textit{Label} (9), \textit{Feedback} (11), and \textit{Navigation} (21).

\add{We treat each manually identified \func error as a ground-truth case and compare tool reports against this set.
In our analysis, we computed performance metrics, such as precision and recall, with respect to app-level \func errors only.
For \toolName, interaction traces that the Accessibility Analyzer classified as agent or infrastructure failures were not reported as accessibility errors and therefore not considered as positive detections.
}

\subsubsection{Results}
Table~\ref{tab:checker_performance} shows performance metrics for the accessibility checkers evaluated on the \func error dataset.
As expected, Accessibility Scanner and ScreenAudit only detected a very limited set of \func errors (4 and 10, respectively) because they did not capture or analyze the interactions with screen readers.
We did not calculate their precision, in part because they reported other aspects of accessibility issues not covered by our dataset, and in part because they were not designed to analyze \func errors \add{in the first place}.

\textbf{Groundhog.}
It detected 20 \func errors (54.1\%) correctly out of a total of 37 in the Locatability and Actionability categories, which corresponds to an overall recall of 25.6\%.
Unfortunately, in these real-world apps, Groundhog demonstrated low precision (14.2\%).
Table~\ref{tab:outcome_breakdown} shows the reasons why many false positives were observed.
For Locatability errors, 46 of the 63 identified elements (73.0\%) were false positives, mainly due to inconsistent or faulty view hierarchy information supplied by the app.
For Actionability errors, 72 of the 75 identified elements (96.0\%) were false positives.
The main reason was the challenge when comparing the resulting screens using touch and \add{a} screen reader (e.g., different dynamic contents or ads, inconsistent banners or pop-ups).
We illustrate some of these examples in Figure~\ref{fig:gh_fail}.

As Groundhog did not report all elements analyzed, \add{(and thus we do not know the true negatives)}, and \del{that}{because} it was not designed to detect all \func errors, we did not attempt to calculate its false positive rate (FPR).
However, given Groundhog's significantly higher number of false positives (118) than that of \toolName (23) when operated on the same dataset, we expect its FPR to also be higher.

\textbf{\toolName.}
\toolName generated and executed 292 tasks. Out of the 78 \func errors, it correctly identified 48 (61.5\%), with a precision of 66.2\%.
\toolName correctly marked 212 (72.6\%) of all tasks as passing with no accessibility issues.
Table~\ref{tab:outcome_breakdown} details the causes for \toolName's false positives and false negatives.
\toolName generated 23 false positive detections ($FPR=9.8\%$).
Of these, seven were due to agent execution issues.
One common problem was navigating the interface inaccurately due to LLM misunderstanding.
Task generation failures accounted for another 16 instances. The majority (12) stemmed from errors in visual UI understanding, including OCR errors in two cases.
Another four \add{problems} were caused by incorrect contextual understandings of non-clickable elements, such as advertisements within apps.
We illustrate some of these examples in Figure~\ref{fig:ta_fail}.

\toolName missed 30 \func errors in the dataset.
These \add{omissions} were caused by the Task Generator failing to generate a relevant task (21 cases), the Accessibility Analyzer not detecting an error (6 cases, including the 2 redundant controls), and the agent using an alternative, accessible interaction sequence which bypassed the error (3 cases).

\add{
When comparing the results, Groundhog achieves higher recall than \toolName within Locatability, but it does so by reporting substantially more false positives, resulting in low precision.
In contrast, \toolName reports fewer total errors, but with much higher precision, and it additionally detects categories that Groundhog cannot represent.
}

\textbf{Cost.}
Groundhog spent an average of 766 seconds ($sd=872$\,s) analyzing each screen.
\toolName spent an average of 1129 seconds ($sd=904$\,s) per screen, consuming 280\,k input tokens and 16\,k output tokens.
At the \edit{current}{August 2025} GPT-4o pricing, \edit{this}{the cost} amounts to about US\$\,0.61 per screen.

\section{Qualitative Evaluation}

To further understand the \textit{types} of \func errors detected, we performed qualitative analysis on the experiment results.
Our analysis followed guidelines set forth in the WCAG 2.2~\cite{w3c_wcag2.2} and the WCAG-EM 1.0~\cite{w3c_wcag_em}.
We also cross-referenced the expert labels for functionality errors~\cite{zhong2025screenaudit} to ensure consistency.

As an overview, we summarized the \func error coverage for accessibility checkers in Table~\ref{tab:checker_coverage}, where we also included AXNav~\cite{taeb24axnav} for reference.
The coverage indications do not mean the checkers can detect \textit{all} errors in a category, but at least \textit{some} \edit{instances}{errors}.



\begin{table}
    \footnotesize
    \centering
    \caption{\add{Detection coverage of different accessibility checkers for each \func error category.}}
    \label{tab:checker_coverage}
    \renewcommand{\arraystretch}{1.5}
    \begin{tabular}{l c c c c c}
        \toprule
        & \makecell[c]{\textbf{Loca-}\\\textbf{tability}}
        & \makecell[c]{\textbf{Action-}\\\textbf{ability}}
        & \textbf{Label}
        & \textbf{Feedback}
        & \textbf{Navigation} \\
        \midrule
        \rowcolor{gray!10}
        \makecell[l]{\textbf{Accessibility}\\\textbf{Scanner}~\cite{a11y_scanner}}
            & --- & --- & \(\bigcirc\) & --- & --- \\

        \textbf{ScreenAudit}~\cite{zhong2025screenaudit}
            & --- & --- & \(\bigcirc\) & --- & \(\bigcirc\) \\

        \rowcolor{gray!10}
        \textbf{Groundhog}~\cite{Salehnamadi2023groundhog}
            & \(\bigcirc\) & \(\bigcirc\) & --- & --- & --- \\

        \textbf{AXNav}*~\cite{taeb24axnav}
            & \(\bigcirc\) & \(\bigcirc\) & --- & --- & \(\bigcirc\) \\

        \rowcolor{gray!10}
        \textbf{TaskAudit} (ours)
            & \(\bigcirc\) & \(\bigcirc\) & \(\bigcirc\) & \(\bigcirc\) & \(\bigcirc\) \\
        \bottomrule
    \end{tabular}
    \bigskip

    \footnotesize \parbox{8cm}{\(\bigcirc\) indicates that the tool detected at least some errors in the category, or claims to support detection. \\ \-\,--- indicates the tool did not detect errors of this type in our evaluation and does not claim to support their detection. \\ \-\:* \,\:We did not evaluate AXNav as it is not publicly available. Coverage is inferred from its system description in~\cite{taeb24axnav}.}
\end{table}

\begin{figure}
  \includegraphics[width=0.473\textwidth]{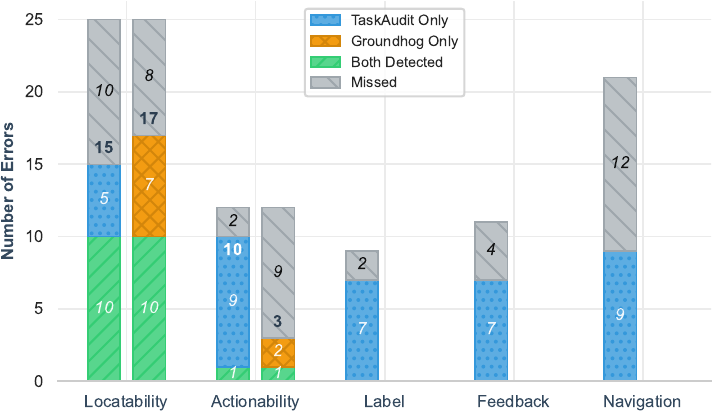}
  \caption{Errors detected by Groundhog and \toolName by category. Bold numbers indicate total errors detected. Groundhog does not support the Label, Feedback, and Navigation categories, and therefore these are not included in the chart.}
  \label{fig:error-breakdown}
  \Description{A bar chart that splits the functionality errors detected by Groundhog and TaskAudit into five categories. For the Locatability category, which had 25 errors, 10 were detected by both tools, 5 were detected by TaskAudit only, 7 by Groundhog only, and 3 were missed by both. In the Actionability category with 12 errors, 1 was detected by both, 8 by TaskAudit only, and 2 by Groundhog only. For the Label category, which had 9 errors, 7 were detected by TaskAudit only, and 2 were missed by both. In the Feedback category with 11 errors, 7 were detected by TaskAudit only, and 4 were missed by both. Lastly, for the Navigation category with 21 errors, 9 were detected by TaskAudit only, and 12 were missed by both.}
\end{figure}

\begin{figure*}
  \includegraphics[width=0.65\textwidth]{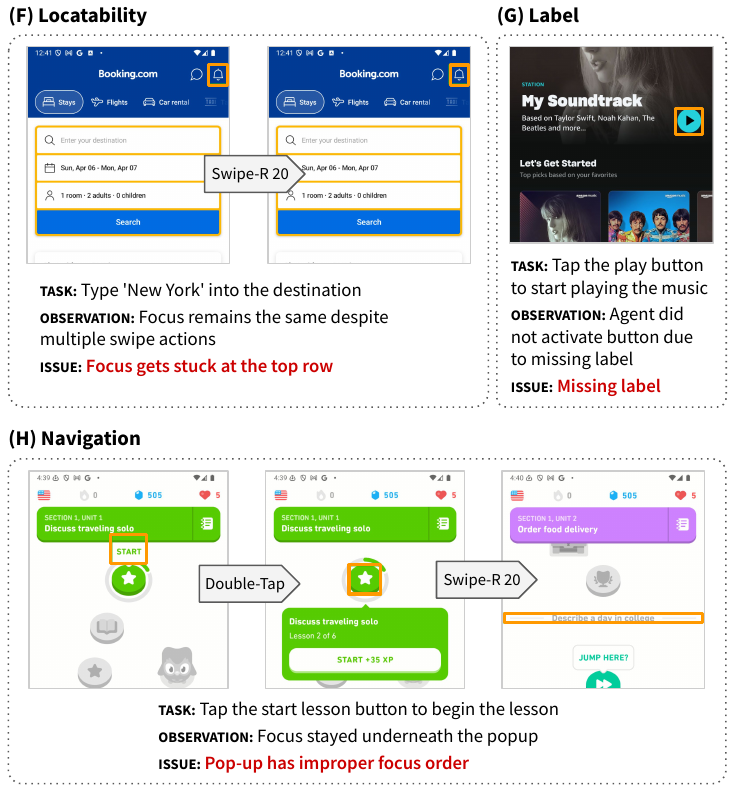}
  \caption{Additional \func errors \toolName detected\add{, with their task descriptions (\textsc{task}), agent observations (\textsc{observation}), identified issues (\textsc{issue}), and elements of focus (solid outline)}. (Continued from Figure~\ref{fig:exp1}.)}
  \label{fig:exp2}
  \Description{Three additional examples of functionality errors detected by TaskAudit. Part (f) shows a locatability error where the screen reader's focus gets stuck in the top row. Part (g) shows a label error where a play button is missing a text label. Part (h) shows a navigation error where a pop-up has an improper focus order.}
\end{figure*}

Figure~\ref{fig:error-breakdown} shows a breakdown of \func errors detected by Groundhog and \toolName.
The two tools demonstrate complementary capabilities within Locatability and Actionability (the only categories Groundhog supports).
In these areas, Groundhog detected several errors missed by \toolName (7 Locatability, 2 Actionability).
\toolName additionally covers Label, Feedback, and Navigation errors, which fall outside of Groundhog's scope.
For missed detections in these categories, most occurred in Navigation (12 cases), primarily because the Task Generator did not produce a relevant task (6 cases), with 3 additional cases missed by the Accessibility Analyzer and 3 cases where the agent performed an alternative accessible sequence.
For Feedback errors, 1 case was missed due to the Task Generator and 3 because the Accessibility Analyzer failed to detect the issue.
For Label errors, all 2 cases were due to the Task Generator not generating the relevant tasks.

Figure~\ref{fig:exp1} and Figure~\ref{fig:exp2} show eight exemplar \func errors detected by \toolName, which we classify into our five error categories.
Two examples represent Locatability errors: a swap-direction button that is not focusable at all (A), and an interface where the screen reader's focus becomes trapped (F).
Example (B) is an Actionability error, where an input field can be focused but not activated.
For Label errors, we identified a direct \textit{label-functionality mismatch} where a "search" tab was mislabeled as "explore" (C), as well as a play button with a missing label that made it undiscoverable (G).
Example (D) highlights an issue with \textit{inappropriate feedback}: a drop-down menu that changed visually but gave no auditory confirmation.
Finally, we classify two examples as \textit{cluttered navigation}: an instance caused by an excessive number of focusable elements (E), and a case where activating a button leads to a pop-up where the subsequent elements do not have proper focus order (H).

\section{Discussion}
    \toolName~presents a novel approach of evaluating mobile app accessibility enabled by task-executing agents.
In this section, we discuss \toolName's implications, limitations, and opportunities for future work.

\subsection{Implications on Automated Accessibility Testing}
\toolName~demonstrates the feasibility of implementing automated testing of \func errors for a variety of error types.
To our knowledge, \toolName is the first automated tool capable of detecting several types of errors previously missed by accessibility checkers: label-functionality mismatch, inappropriate feedback, and cluttered navigation.

Our key insight is the direct simulation of screen reader user experience using multi-agent task execution to uncover errors.
We expand on prior work such as Groundhog~\cite{Salehnamadi2023groundhog} by integrating agentic execution, reflection, and semantic understanding.
We show that our multi-agent design achieves high reliability in executing tasks via the screen reader proxy.
This agentic method is particularly effective at detecting error types requiring contextual understanding, such as inappropriate feedback, label-functionality mismatches, and navigation issues, as these depend on assessing an action's outcome against its intended purpose.
We also show that LLM-supported accessibility auditing~\cite{zhong2025screenaudit} can be transferred to examine \func errors from automated interaction traces and detect multiple types of errors.

\subsubsection{\add{Complementing Existing Methods}}\label{sec:implications_complementing}
Our results also highlight the complementary strengths of different automated methods.
While \toolName's semantic analysis allows it to detect entire categories of errors missed by other tools, Groundhog's structured crawling identified several Locatability and Actionability errors that our agent-based approach did not (Figure~\ref{fig:error-breakdown}).
This suggests that a future hybrid approach, combining agentic semantic analysis with exhaustive mechanical crawling, could yield more comprehensive accessibility audits.
\add{For example, accessibility metadata inspection can be combined with Task Generator outputs to filter out invalid controls (e.g. those that are occluded or hidden) and flag potential Locatability and Actionability errors, which can then be quickly evaluated with a method like Groundhog.
This would allow \toolName to focus on addressing elements with no apparent Locatability and Actionability errors, which would reduce the amount of agentic execution needed, improving efficiency and reducing cost.
}

\add{
On a larger scale, ecosystem-level analyses~\cite{ross_epidemiology_taccess20, fok_large-scale_chi22, yan_current_state_accessibility, chen2022accessible} identify common accessibility failures and quantify their prevalence across large app populations.
\toolName complements these methods by providing task-based functionality analysis on a smaller set of critical flows.
A practical pipeline might first apply large-scale screening to locate high-risk apps and screens, then deploy TaskAudit as a targeted check on those subsets to evaluate \func errors.
}

\subsubsection{\add{Extensibility}}\label{sec:implications_extensibility}
\add{We designed \toolName's architecture to be modular. 
The Task Generator, Task Executor, and Accessibility Analyzer communicate through structured representations.
For example, different LLMs or prompting strategies can be used for task generation and alternative agent frameworks can be used for execution.
The screen reader proxy can also be replaced with other assistive technology proxies, such as voice-based interfaces or switch access systems.
}

\add{
At the same time, \toolName is different from many recent computer-use~\cite{wang2025opencua, openai2025cua} and mobile agents~\cite{wang2024mobileagentv2, zhang2023appagent} that operate primarily over screenshots.
Our agents do not observe the raw pixels, but act on TalkBack outputs, and must reason about the UI from the perspective of a screen reader user.
This interface makes the system less dependent on a model's vision capabilities, but more dependent on its ability to follow instructions and maintain context over an interaction trace.
We see an opportunity to fine-tune smaller language models that learn to operate screen readers directly, potentially via reinforcement learning on TalkBack transcripts and action traces.
This could enable future agents to execute accessibility tasks more reliably and at lower cost.
}

\subsubsection{\add{Adapting to Other Interaction Modalities}}
The effectiveness of \toolName in detecting Feedback and Navigation errors demonstrates the value of analyzing interaction sequences from a non-visual, transcript-based perspective.
This has implications beyond screen readers\edit{, suggesting that}{.} Similar agentic methods could be adapted to test other non-traditional interaction modalities, such as voice-based interfaces or switch access systems, where the correctness of an interaction depends \del{entirely }on the sequence and quality of non-visual feedback.
\add{For example, voice-based interfaces can be evaluated by simulating voice commands, navigating through voice-based menus, and analyzing spoken feedback transcripts.
Switch access can be tested by simulating sequential inputs and analyzing the navigation order, focusability and actionability of elements.}

\subsubsection{\add{Integrating into Development Workflows}}
\add{
In practice, accessibility is typically checked in various stages during software development.
Teams run automated rule-based tools inside their IDEs and continuous integration (CI) pipelines~\cite{axe,deque2024axelint}, complement them with runtime checkers like the Accessibility Scanner~\cite{a11y_scanner}, while also rely on manual assistive technology walkthroughs using tools such as TalkBack~\cite{talkback,pellegrini2020prioritize}.
}

\add{
We envision \toolName fitting into this workflow by integrating with existing accessibility inspection tools and IDEs.
It can also be automatically triggered when screens or interaction flows change, alongside existing accessibility checks.
Similar to how code quality gates~\cite{schermann2016qualitygates} and security testing~\cite{potter2004software} can block deployment when CI pipelines fail key criteria, \toolName can be included into regression test suites or CI pipelines for high-impact flows.
For example, organizational policy can require a feature to launch only when its \func audit report has been reviewed and reproducible \func errors are resolved.
}

\add{
Although task-based audits introduce additional up-front costs (i.e., primarily in LLM calls), catching \func errors earlier can reduce expensive last-minute fixes.
By surfacing problems before user testing, \toolName can reduce engineering and QA effort spent resolving accessibility issues that require redesigns.
}

\add{
Practically, running task-based audits requires infrastructure to deploy apps, manage emulators, and integration into existing test environments, which can increase both compute and engineering overhead.
We note that similar infrastructure is already common in modern development workflows for end-to-end testing~\cite{android_monkey, aws_farm, appium}, and prior work in accessibility evaluation will require similar engineering efforts in deployment~\cite{taeb24axnav, zhong2025screenaudit}.
}

\subsection{Limitations}
\toolName currently cannot perform touch exploration, limiting its ability to detect certain errors that depend on spatial interaction patterns common among screen reader users.
This constraint means errors that appear exclusively when users tap directly on elements might remain undiscovered.
The system's effectiveness is also bounded by the capabilities of its components.
Task generation relies on visual UI understanding; related failures, such as OCR errors or misinterpreting elements, occurred in our evaluation and can limit interaction coverage.
Furthermore, the quality of the generated task prompts remains a crucial LLM-dependent limitation.
Poor or overly simplistic task generation can lead to limited coverage of real-world usage scenarios, which contributed to missed errors in our evaluation.
\add{Of the 30 false negatives in \toolName, 21 can be attributed to the Task Generator. 
A potential way to increase the diversity of generated tasks is to run the generator multiple times and select the ``best'' tasks (e.g., according to their coverage and detail levels).
}

A limitation we observed during task execution is that our reflection agent can sometimes be less effective because it relies solely on the next screen's initial transcript to assess task progress.
As this accurately reflects a screen reader user's experience, it can make the agent miss visually available cues that indicate task failure or success, resulting in unnecessary explorations.
Task execution agents also encountered issues, sometimes struggling with inconsistent interface behavior, string matching limitations, overly verbose screen reader transcripts, or proxy reliability, resulting in false positives or overlooked errors.

Additionally, the system currently cannot detect errors involving functionality redundancy, such as when multiple controls perform the same action.
While we adopted one effective agent setup, we have not systematically compared alternative language models, agent architectures, or combinations thereof.
Different groupings of models and agents may yield varying efficiencies or accuracies that we do not currently capture.

\add{Finally, \toolName currently incurs computational costs due to multiple LLM calls for task generation, execution, and analysis.
The time spent per screen is also substantially longer than Groundhog.
We anticipate that combining \toolName with rule-based analysis (see Section~\ref{sec:implications_complementing}) can help reduce the number of screens requiring agentic evaluation, thereby lowering overall costs.
Model optimization, such as hosting smaller, fine-tuned models locally (see Section~\ref{sec:implications_extensibility}) can also help improve efficiency and reduce costs.
}

\subsection{Future Work}

\add{While we proposed the concept of ``\func error,'' we did not attempt to construct a complete taxonomy of these errors.
Instead, we relied on prior work and qualitative observations during our experiments to identify five error categories.
\toolName provides a useful technical framework that can potentially speed up the identification of new error instances.
This will enable future research to explore whether additional error types exist, their impact on users, and practical solutions to addressing them.
}

\del{
Future extensions of our work should explore integrating structured crawling approaches, similar to Groundhog~\cite{Salehnamadi2023groundhog}, to systematically enhance UI coverage.
Such crawler-integrated agents could achieve more representative user behaviors by actively prioritizing unexplored UI interactions.
}

\add{We previously discussed optimization opportunity in execution, where  fine-tuning smaller language models directly on TalkBack interaction can potentially lead to reliability and cost improvements.}
In parallel, incorporating screenshot analysis can further enhance reflection-agent decisions.
In cases where it is visually obvious that a control cannot be focused or activated, \del{the agent will still attempt to explore the screen.
This can be improved by }augmenting the agent with vision models when assessing interaction outcomes\edit{, allowing for the early termination of}{ could terminate} obviously impossible tasks and \edit{improving}{improve} testing efficiency~\cite{lu2024omniparser, baechler2024screenai, wang2021screen2words}.

Another potential direction is to automatically suggest actionable solutions \edit{in addition to}{after} detecting accessibility errors.
Future research may leverage code-analysis methods to highlight relevant code segments causing these issues, along with tailored recommendations for developers.
Such integration could ease remediation and improve mobile app accessibility.

\section{Conclusion}
    \add{In this paper, we introduced the concept of} ``\func errors'' \add{that} are often overlooked in accessibility evaluation and quality assurance due to their complexity and lack of tool support.
\toolName expands current automated accessibility error checking by focusing on interactive tasks and automating the evaluation of these tasks using a multi-agent approach.
Our results show the potential of agent-driven accessibility evaluation and simulated user testing in identifying accessibility errors.
By enabling earlier and more comprehensive detection of \func issues, \toolName can support app developers and QA professionals in creating more inclusive mobile experiences.
Future work can explore the integration of rule-based checkers and programmatic crawlers to improve error detection rate.
Opportunities also exist for conducting code-level analysis and generating remedies to remove the errors.


\begin{acks}
This work was supported in part by a \grantsponsor{google}{Google Research Award}{} and by an \grantsponsor{amazon}{AWS Agentic AI Amazon Research Award}{}.
Any opinions, findings, conclusions or recommendations expressed in our work are those of the authors and do not necessarily reflect those of any supporter.
\end{acks}

\bibliographystyle{ACM-Reference-Format}
\bibliography{ref}

\appendix

\section{Prompts Used in \toolName}
    \label{sec:appendix-prompt}
    \subsection{Decision}
The following is the prompt used in the Decision Agent.

\scriptsize
\begin{lstlisting}
    You are a mobile agent that performs tasks using a screen reader. By performing these tasks, we aim to identify potential accessibility issues in the app.
    ### Background ###
    - You are on an Android phone.
    - You are using a screen reader to interact with the phone.
    - You are provided with transcripts of the screen reader's output for your last action.
    - If you are given no transcript, try to swipe {direction} 20 times to explore the screen. If you still cannot find the it, try swiping {direction} 20 time for at most {count} times to explore the screen before you conclude the task is impossible.
    - If a transcript item is <wrap>, it means the swiping gesture has reached the end of the screen and wrapped around to the other end.
    - The user's instruction/task is: <|begin_task|> {task} <|end_task|>
    
    ### Screen Reader Transcript Information ###
    In the screen reader transcript, the transcripts are in the following format:
    ```
      {{
        "index": <index of the element>,
        "transcript": "<transcript of the element>"
      }},
    ```
    The screen reader transcripts for the last action are:
    <|begin_transcript|>
    {transcript_dict}
    <|end_transcript|>
    You are currently focused on the LAST element of the transcript list. Which is: 
    <|begin_focused_element_transcript|>
    {last_element_transcript}.
    <|end_focused_element_transcript|>
    
    Notice:
    - The transcript list is a result of your previous action.
    - The length of the list depends on your action(s) and the last element of the list may not be the last element of the whole screen.
    
    ### Keyboard Status ###
    We extract the keyboard status of the current screenshot about whether the keyboard is activated.
    The keyboard status is: {keyboard_status_str}
    
    {additional_info_str}

    ### Task ###
    # Propose a *single* action to be performed on the current state.
      ## You should not output a list of actions to be performed consecutively. Instead, you must output possible immediate actions that can be performed on the current state.
      ## If you don't know what to do, propose something that would help explore the environment and gather more information to solve the problem.
      ## If no talkback transcript provided nor any context, you can can propose something like {{"action_type": "{action_direction}", "repetitions": "20"}} to loop through the elements and collect info from the talkbacks.
      ## Please judge if the user's instruction has already been completed based on ### Progress ###. If so, pick 'STATUS_TASK_COMPLETE' as 'action_type'.
      ## Think step by step and provide every detail.
 
    ### Output Format ###
    # Step 1: Put your thought process within `<|begin_think|>` and `<|end_think|>`.
    # Step 2: Describe what you propose to do next enclosed with `<|begin_action|>` and `<|end_action|>`. 
      Also please generate a brief natural language description for each action based on your thought.
      Represent all actions you propose in a list of dict literal:
      ```
      [{{"action": <action_json>, "description": <description_of_action>}}]
      ```
      For the <action_json>, there are only the following types of actions to choose from:
        1. {{"action_type": "SWIPE_RIGHT", "repetitions": "<count>", "stop_at": "<element_transcript>", "stop_at_occurrence": "<occurrence>"}}
        2. {{"action_type": "SWIPE_LEFT", "repetitions": "<count>", "stop_at": "<element_transcript>", "stop_at_occurrence": "<occurrence>"}}
        3. {{"action_type": "DOUBLE_TAP"}}
        4. {{"action_type": "PRESS_BACK"}}
        5. {{"action_type": "TYPE", "typed_text": "<text>"}}
        6. {{"action_type": "WAIT"}}
        7. {{"action_type": "STATUS_TASK_COMPLETE"}}
        8. {{"action_type": "TASK_IMPOSSIBLE"}}
  
    ### Hints: ###
      - If you want to type something, please check if the keyboard is visible (the input box is focused). If not, you need to TOUCH the input box first.
      {typing_additional_str}
      - Use SWIPE_LEFT and SWIPE_RIGHT to explore elements. They will move current focus one by one in the direction specified. SWIPE_LEFT means moving to the previous element, SWIPE_RIGHT means moving to the next element. Use "repetition" to indicate the number of elements to move.
      - When you are exploring, set "stop_at" to "".
      - When you would like to focus on a specific element, set "stop_at" to the full transcript of that element. There may be multiple items of the same transcript; use "stop_at_occurrence" to indicate which occurrences to stop at (e.g. 1 means stop at first time we see it; 2 means skip the first one). Also set "repetition" to the number needed to find the element.
      - DOUBLE_TAP means activating the focused element. You must first check if the focused element in <focused_element_transcript> is desired, then activate it to interact with it.
      - Choose STATUS_TASK_COMPLETE when you think the task has been completed according to the progress made and the final transcript.
      - Choose TASK_IMPOSSIBLE when you think the task is impossible to complete.
      - Always use double quotes for JSON keys and string values.
\end{lstlisting}
\normalsize

\subsection{Reflection}
The following is the prompt used in the Reflection Agent.

\scriptsize
\begin{lstlisting}
    You are an Android mobile agent that assists a user to accomplish tasks using a screen reader.
    ### Progress before the current operation ###
    {progress_str}
    The keyboard status is: {keyboard_status_str_1}
    
    ### Transcript after the current operation ###
    The screen reader transcripts after the last action are:
    <|begin_transcript|>
    {transcript}
    <|end_transcript|>
    The keyboard status is: {keyboard_status_str_2}
    
    ### Current operation ###
    The user's instruction is: {task}. 
    In the process of completing the requirements of instruction, this operation is performed on the phone. Below are the details of this operation:
    # Operation thought: {thought}
    # Operation action: {action}
    
    ### Response requirements ###
    Now you need to output the following content based on the screenshots before and after the current operation:
    Whether the result of the "Operation action" meets your expectation of "Operation thought"?
    A: The result of the "Operation action" meets my expectation of "Operation thought".
    B: The "Operation action" results in a wrong page and I need to return to the previous page.
    C: The "Operation action" produces no changes.
    
    ### Hint ###
    # You should not judge based on if this single operation can complete the user's instruction. Instead you should judge based on if this operation made any progress towards the goal.
      For example, if the user's task is to search something and the operation is to open the browser:
      You should not expect this browser to show the search page directly. As long as the browser is opened, you should judge this operation to be A.
    # You should also encourage exploration and gathering more information to solve the problem.
    # If the screen reader transcript is empty, it means the screen reader has not output any information. You should not judge this operation to be C.
    
    ### Output format ###
    Your output should be strictly a dict literal with the following format:
    ```
    <|begin_think|>
    Your thought for this.
    <|end_think|>
    <|begin_answer|>
    Answer of the above question, A or B or C.
    <|end_answer|>
    ```
\end{lstlisting}
\normalsize

\subsection{Accessibility Analyzer} \label{sec:appendix-analyzer}
The following is the prompt used in Stage 1 of the Accessibility Analyzer.

\scriptsize
\begin{lstlisting}
    Your task is to carefully analyze the provided action trace from a mobile device screen reader to determine its outcome. Pay close attention to the details.
    Your analysis must consider two independent aspects:
    1. Overall Task Status: Check if the 'Success Criterion' from the overall Task Specification has been met in the 'after' transcript. A task can be completed at any step.
    2. Immediate Action Status: Check if the specific action just taken was successful on its own, based on the provided action-specific heuristics.
    Action-Specific Heuristics:
    - For a 'double-tap' on a button/menu, new elements or a screen change are expected.
    - For a 'double-tap' on a text input, a keyboard or 'editing mode' is expected.
    - For a 'swipe' or 'scroll', the transcript must change.
    - For 'type' actions, the typed text should be announced.
    Respond ONLY with a JSON object in the following format.
    {
      "overall_task_status": "COMPLETE or INCOMPLETE",
      "task_status_reasoning": "A concise, one-sentence explanation for the overall task status, based on the success criterion.",
      "immediate_action_status": "SUCCESS or FAILURE",
      "action_status_reasoning": "A concise, one-sentence explanation for the immediate action's status, based on the heuristics."
    }
    
    Task Specification:
    - DESC: {description}
    - ELEM: {element}
    - CRIT: {criterion}
    
    Action Taken: {action_taken}
    
    Screen Reader Transcript BEFORE Action:
    <|begin_transcript|>
    {transcript_before}
    <|end_transcript|>
    
    Screen Reader Transcript AFTER Action:
    <|begin_transcript|>
    {transcript_after}
    <|end_transcript|>
\end{lstlisting}
\normalsize

The following is the prompt used in Stage 2 of the Accessibility Analyzer.

\scriptsize
\begin{lstlisting}
    An action performed by an automated agent has failed. Your task is to perform a detailed root-cause analysis to determine if this failure was caused by a specific accessibility error. Please be meticulous in your analysis and use the provided transcripts as evidence.
    Respond ONLY with a JSON object in the following format. The 'thought' field is for your step-by-step reasoning.
    {
      "thought": "A step-by-step analysis of the action trace. First, state the intended action. Second, describe the actual outcome from the transcripts. Third, pinpoint the discrepancy. Finally, reason whether this discrepancy constitutes an accessibility error.",
      "problematic_element": "The name of the UI element from the transcript that caused the issue. If not applicable, use 'N/A'.",
      "element_index": int,
      "explanation": "A clear, evidence-based explanation for your conclusion. If an accessibility error occurred, explain how it fits the chosen category. If no accessibility error occurred, explain the likely non-accessibility cause (e.g., agent mistake, unexpected app state change)."
    }
    
    Task Specification:
    - DESC: {description}
    - ELEM: {element}
    - CRIT: {criterion}
    
    Failed Action: {action_taken}
    Failure Reason from Stage 1: {stage_1_reasoning}
    
    Screen Reader Transcript BEFORE Action:
    <|begin_transcript|>
    {transcript_before}
    <|end_transcript|>
    
    Screen Reader Transcript AFTER Action:
    <|begin_transcript|>
    {transcript_after}
    <|end_transcript|>
\end{lstlisting}
\normalsize










\end{document}